\documentclass[
    ,final            
  ]
  {aipproc}

\layoutstyle{6x9}

\renewcommand\XFMtitleblock{
  \XFMtitle
  \let\XFMoldpar\par
  \def\par{\XFMoldpar\def\par{\space
	for the \CTA\ Consortium\XFMoldpar}}
   \XFMauthors
   \let\par\XFMoldpar
   \XFMaddresses
   \XFMabstract
   \vspace{5pt}
   \XFMkeywords
   \XFMclassification
}

\newcommand{\CTA}{CTA}
\newcommand{\HESS}{H.E.S.S.}
\newcommand{\BESS}{\emph{BESS}}

\begin{document}

\title{Trigger and data rates expected for the \CTA\ Observatory}


\author{Manuel {Paz Arribas}}{
  address={Institut f\"ur Physik, Humboldt-Universit\"at zu Berlin, Newtonstr. 15, D-12489 Berlin, Germany},
  altaddress={DESY, Platanenallee 6, D-15738 Zeuthen, Germany},
  email={mapaz@physik.hu-berlin.de} 
}

\author{Ullrich {Schwanke}}{
  address={Institut f\"ur Physik, Humboldt-Universit\"at zu Berlin, Newtonstr. 15, D-12489 Berlin, Germany},
}

\author{Ralf {Wischnewski}}{
  address={DESY, Platanenallee 6, D-15738 Zeuthen, Germany},
}

\begin{abstract}
The Cherenkov Telescope Array (\CTA) is an initiative to build a next-generation observatory for very-high energy $\gamma$-rays. Its expected large effective area ($\mathcal{O}(10^{7}\mathrm{m}^2)$) and energy threshold as low as 25 GeV imply a challenge for triggering and data acquisition systems. The analysis of the official \CTA\ Monte Carlo production-1 simulations leads to array trigger rates of $\mathcal{O}$(10 kHz) and data rates ranging from $\mathcal{O}$(100 MB/s) to $\mathcal{O}$(1000 MB/s), depending on the read-out scenario.
\end{abstract}

\maketitle


\section{Trigger simulations}

The estimation of the array trigger rates, data rates and data volumes is important for the design of the data acquisition and archival system for \CTA. In order to perform these estimations, some input is needed regarding the camera electronics and the information that should be stored. Previous studies for the \CTA-97 benchmark array are discussed in \cite{thesis:diplomarbeit}, where the methods used for the study in this section are discussed in more detail.

In the following, a detailed study of the trigger rates, data rates and data volume for the different candidate arrays (southern site: A-K, northern site: NA, NB) of the \CTA\ Monte Carlo production-1 is presented.
Each of the \CTA\ sites will consist of an array of telescopes of different types: large, medium and small size telescopes (LST, MST and SST, respectively). The trigger rates of single telescopes of each type are also presented.
A subset of four MSTs organized as a \HESS-like telescope array (designated as subarray HESS) is also studied for comparison purposes. Such a subarray is also interesting for a possible operation mode, where the \CTA\ array would be split into subarrays of telescopes, each one observing one region of the sky. More details on the simulated data sets and the candidate arrays used for the analysis are presented in \cite{paper:cta_mc}.

In this study, a three level trigger was simulated: two levels (0 and 1) for the camera and one level (2) for the array trigger.
For the camera trigger, a majority trigger was simulated, where at least one pixel and two of its next-neighbors (level 1) should all have passed individual thresholds of four photo-electrons (level 0).
The duration of the allowed gate for pixel coincidence for each telescope type is 3 ns for LSTs, 6 ns for MSTs and 16 ns for SSTs. The camera trigger of each telescope type is adjusted accordingly to yield less than 100 Hz random triggers per telescope due to night sky background (NSB). For the array trigger, a simple stereo trigger was simulated, where at least two telescopes out of the whole array should send a camera trigger (level 2). In the case of the candidate array E, this trigger was compared to a more complex one, where the topology of the array is taken into account and only stereoscopy among next-neighboring telescopes is allowed.

\section{Trigger rates}

To obtain the expected trigger rate, the trigger effective areas need to be folded with an appropriate particle spectrum, and then integrated over the energy. In this study, a \HESS-Crab-Nebula-like spectrum \cite{paper:hess_crab} was used for the $\gamma$-ray signal events (for simplicity the cut-off observed by \HESS\ was ignored) and a combination of the proton spectrum measurement of \BESS\ at low energies \cite{paper:bess} and HEGRA at high energies \cite{paper:hegra} is used for the background events. For more details, please refer to \cite{thesis:diplomarbeit}.

The single telescope rates are shown in table \ref{tab:singlerates}. Each of the telescopes will have to cope with trigger rates of $\mathcal{O}$(1 kHz). For the MSTs, a variant of the camera has also been simulated with a large field of view (denoted MST LFoV).

\begin{table}[!htb]
 \centering
 \begin{tabular}{l|c|c|c|c|c}
 tel type & diameter & FoV            & pixel size     & NSB   & rate \\
          & (m)      & ({}$^{\circ}$) & ({}$^{\circ}$) & (MHz) & (kHz)\\
 \hline
 LST      & 23       & 5              & 0.09           & 122   & 4.7  \\
 MST      & 12       & 8              & 0.18           & 120   & 2.2  \\
 SST      & 6.7      & 10             & 0.25           &  85   & 0.9  \\
 MST LFoV & 12       & 10             & 0.27           & 274   & 1.7  \\
 \end{tabular}
 \caption{Camera trigger rates for each of the four different telescope types for a BESS-proton spectrum. The column NSB lists the camera rate of photo-electrons generated by NSB photons at the photo cathode of an individual PMT: these values are adapted to represent the level of NSB under moonless conditions, when pointing away of the Galactic Plane and zodiacal light.}
 \label{tab:singlerates}
\end{table}

By requiring a stereoscopic trigger within one of the candidate arrays (at least two telescopes should trigger), the telescope trigger rates are reduced: in the case of candidate array E to $\sim 2.5$ kHz for LSTs, $\sim 1.5$ kHz for MSTs and $\sim 0.7$ kHz for SSTs.

\begin{figure}[!htb]
  \centering
  \includegraphics[width=3in]{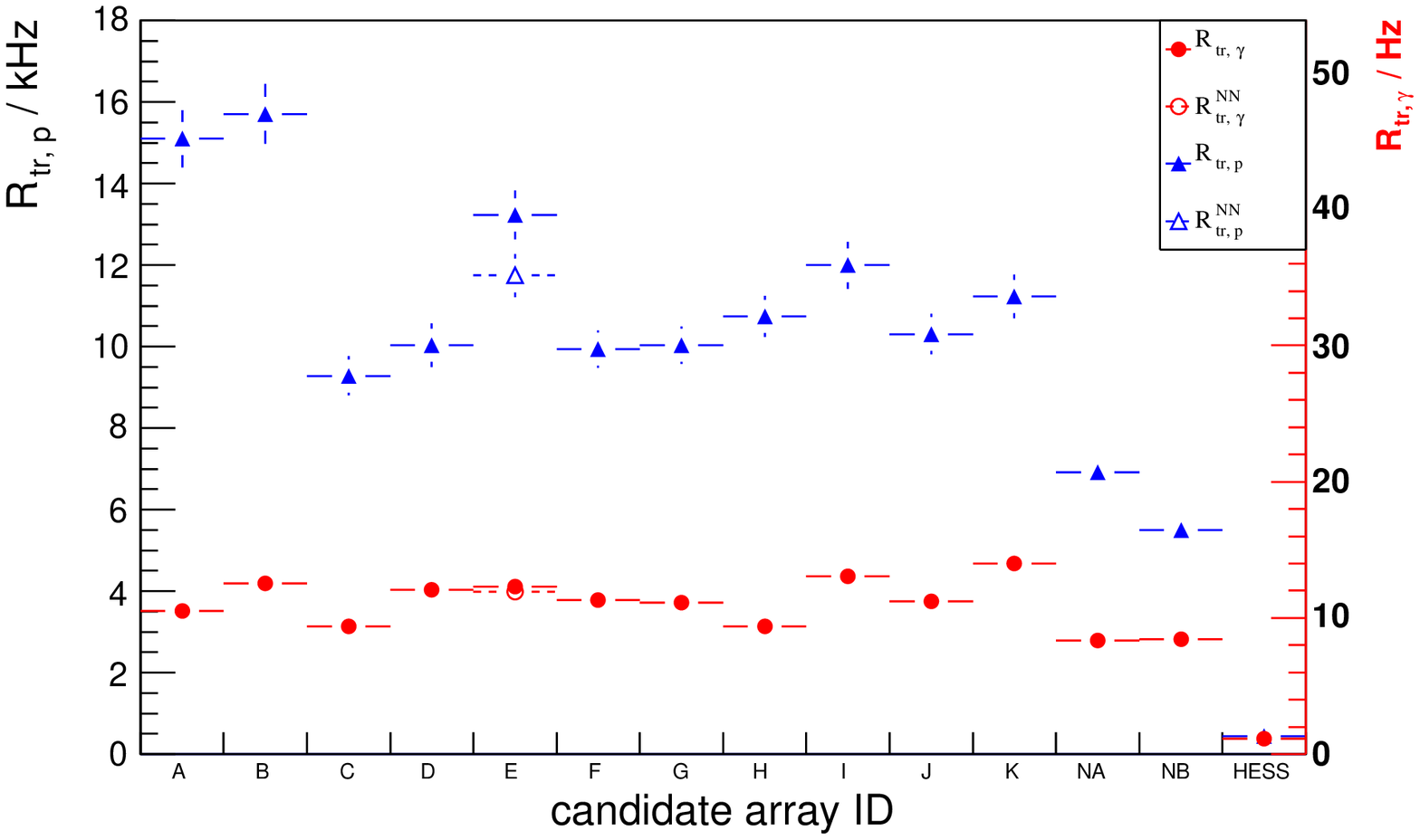}
  \hfil
  \includegraphics[width=3in]{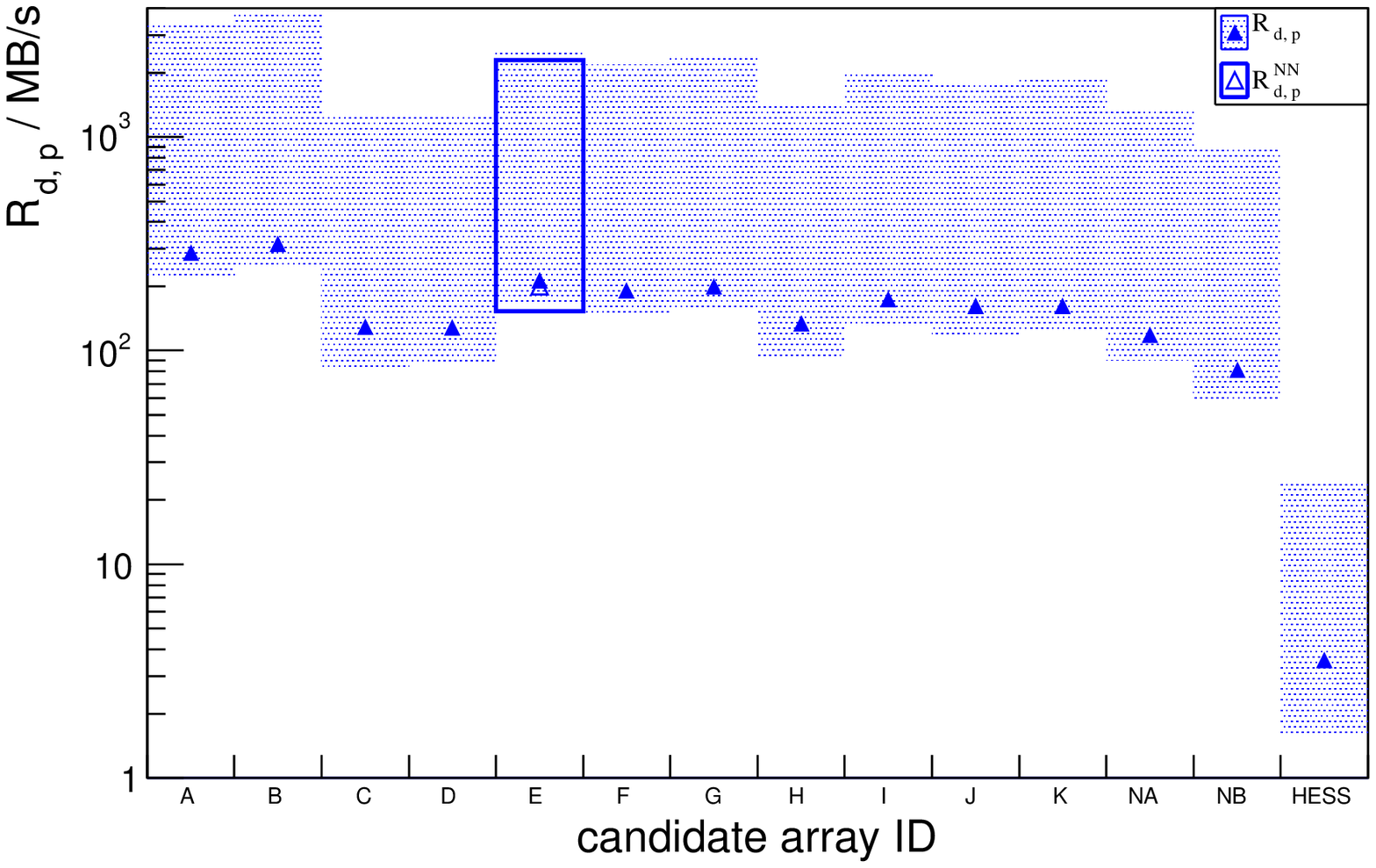}
  \caption{Trigger rates (left) and data rates (right) for all \CTA\ candidate arrays. Left: the blue triangular points represent \BESS-proton trigger rates, and the red circular ones represent 1-Crab-Unit-$\gamma$-ray trigger rates. Note also the different scale for the protons (left scale in kHz) and for the $\gamma$-rays (right scale in Hz). For the case of the candidate array E, dashed open points indicate the values for a next-neighbors triggering system. Right: for each candidate array, the shaded area represents the range of data rate expected from the simplest scenario (lower boundary) to the extreme case (upper boundary). The triangle indicates the value for the intermediate scenario (see the text for details on each scenario). For the case of the candidate array E, an empty box and an open triangle indicate the range of values for a next-neighbors triggering system.}
  \label{fig:CTA_prod1_TR_and_DR}
\end{figure}

The results for each simulated candidate array and for both kinds of particles are shown in figure \ref{fig:CTA_prod1_TR_and_DR} (left). Comparing the trigger rates for $\gamma$-rays and protons for each candidate array, the signal-to-noise ratios are quite similar for all cases ($\mathcal{O}(10^{-3})$): note the different scale for the proton blue triangular points (left scale in kHz) and for the $\gamma$-ray red circular points (right scale in Hz). Only the subarray HESS has a higher signal to noise ratio of $\sim 1/400$, due to its reduced FoV and the fact that the $\gamma$-ray simulations are point-like, whereas the proton ones are diffuse. From the plot it is clear that most of the southern site candidate arrays have a similar array trigger rate between 9 and 13 kHz, only the candidate arrays A and B (tuned to have best performance at low energies) have a higher rate on the order of 15 kHz. The northern site candidate arrays have a lower rate on the order of 6 kHz, and the subarray HESS has a rate of about 400 Hz. It is also interesting that the $\gamma$-ray trigger rates are quite similar for all candidate arrays, except for the subarray HESS.

For the case of the candidate array E, a next-neighbor array trigger was simulated, as mentioned above. This is motivated by the fact that most proton induced showers are very scattered, producing a quite inhomogeneous lightpool at the ground, whereas $\gamma$-ray induced showers tend to be more compact, producing more homogeneous lightpools. Indeed, the trigger rates marked as dashed lines in figure \ref{fig:CTA_prod1_TR_and_DR} (left) show that a 10\% reduction in the proton background is achieved, while keeping most of the $\gamma$-ray signal (only a 3\% loss).

An estimate of the trigger rates expected for the \CTA\ Observatory is given in table \ref{tab:array_data_rates}.

\section{Data rates and data volumes}

Similarly to the trigger rates, the data rates are estimated by folding the effective areas with the flux spectrum and the expected size of the events for each energy bin, before integrating in energy.

For the estimation of the event size, some assumptions have to be made about the amount of data that should be stored per telescope. The simplest scenario would be an integrated signal over a short (15 ns) time window, which translates into two bytes per pixel and channel. The extreme case would be to do waveform sampling for all pixels at a high rate (1 GHz) in that window, in which case 30 bytes per pixel and channel should be stored. An intermediate scenario is also possible, where only pixels important for the reconstruction of the image are read out with sampling, whereas for the rest, the integrated charge is stored.

Assuming PMTs with only one read-out channel (i.e. no separate channels for high and low gain), and that only triggered telescopes are read out, the data rates for the candidate arrays range from $\mathcal{O}$(100 MB/s) for the simplest scenario, to $\mathcal{O}$(1000 MB/s) for the extreme one, as it is indicated by the limits of the shaded areas in figure \ref{fig:CTA_prod1_TR_and_DR} (right). The values for the intermediate case are on the order of a few 100 MB/s, as denoted by the markers on the same plot.
In this case, the next-neighbor array trigger from array E reduces the data rate by 5\%, when compared to the normal stereoscopic trigger.

An estimate of the data rates expected for the \CTA\ Observatory is given in table \ref{tab:array_data_rates}.
Using the candidate array E as an example for the southern site array, the data rate ranges from $\sim 200$ MB/s for the simplest scenario, to $\sim 2500$ MB/s for the extreme one. In the case of northern site candidate array (the candidate array NA is used as example), the values are lower: $\sim 100$ MB/s for the simple scenario and $\sim 1400$ MB/s for the extreme one. The event sizes range from $\sim 20$ kB to $\sim 250$ kB for the southern site candidate array and from $\sim 15$ kB to $\sim 220$ kB for the northern site one. These values, and the corresponding data volumes assuming a 15\% duty cycle are displayed in table \ref{tab:array_data_rates}. The total data volume expected after 15 years of operation amounts to $\sim 20$ PB for the simple scenario and $\sim 260$ PB for the extreme one.

\begin{table}[!htb]
 \centering
 \begin{tabular}{l|c|c|c}
                              & \CTA\ North    & \CTA\ South    & \CTA           \\
 \hline
 trigger rate                 & $\sim 7$\,kHz  & $\sim 13$\,kHz & $\sim 20$\,kHz \\
 event size                   & 15-220\,kB     & 20-250\,kB     & 15-250\,kB     \\
 data rate                    & 100-1400\,MB/s & 200-2500\,MB/s & 300-4000\,MB/s \\
 data volume per month        & 40-520\,TB     & 74-930\,TB     & 120-1500\,TB   \\
 data volume per year         & 0.5-6.1\,PB    & 0.9-11\,PB     & 1.5-20\,PB     \\
 total data volume (15 years) & 6.5-92\,PB     & 13-165\,PB     & 20-260\,PB     \\
 \end{tabular}
 \caption{Estimated trigger rates, event sizes, data rates and data volumes for the \CTA\ observatory. The ranges cover the values from the expectations using the simplest scenario (store only integrated signal) to the extreme one (1 GHz sampling rate for all pixels in the camera). A duty cycle of 15\% is assumed.}
 \label{tab:array_data_rates}
\end{table}

To conclude, we report on the expected trigger and data rates for realistic operation of the \CTA\ Observatory. Our findings are preliminary, with modification likely due to final choice of technology (i.e. camera electronics) and trigger logics. Updated and more detailed estimations are in preparation \cite{thesis:phd_thesis}.


\begin{theacknowledgments}
We gratefully acknowledge support from the agencies and organizations listed in this page: http://www.cta-observatory.org/?q=node/22.
\end{theacknowledgments}



\bibliographystyle{aipproc}   

\bibliography{gamma12_cta_rates}

\IfFileExists{\jobname.bbl}{}
 {\typeout{}
  \typeout{******************************************}
  \typeout{** Please run "bibtex \jobname" to optain}
  \typeout{** the bibliography and then re-run LaTeX}
  \typeout{** twice to fix the references!}
  \typeout{******************************************}
  \typeout{}
 }

\end{document}